\documentclass[autodetect-engine, twocolumn, longbibliography]{revtex4-1}
\usepackage[utf8]{inputenc}
\usepackage{amsmath,amssymb,comment, }

\usepackage{graphicx}
\usepackage{braket, bbm, color, ulem}
\newcommand{\Lind}{\mathcal{L}}

\newcommand{\order}{\mathcal{O}}

\begin{document}

\title{Variational Quantum Algorithm for Non-equilibrium Steady States}
\author{Nobuyuki Yoshioka$^{1,5}$}
\email{\tt nobuyuki.yoshioka@riken.jp}
\author{Yuya O. Nakagawa$^2$}
\author{Kosuke Mitarai$^{2,3}$}
\author{Keisuke Fujii$^{3,4}$}
\affiliation{$^{1}$Department of Physics, University of Tokyo, 7-3-1 Hongo, Bunkyo-ku, Tokyo 113-0033, Japan}
\affiliation{$^{2}$QunaSys Inc.,
    High-tech Hongo Building 1F, 5-25-18 Hongo, Bunkyo, Tokyo 113-0033, Japan.
}
\affiliation{$^{3}$Graduate School of Engineering Science,
    Osaka University,
    1-3 Machikaneyama, Toyonaka, Osaka 560-8531, Japan.
}
\affiliation{$^{4}$JST, PRESTO, 4-1-8 Honcho, Kawaguchi, Saitama 332-0012, Japan.
}
\affiliation{$^{5}$Theoretical Physics Laboratory, RIKEN Cluster for Pioneering Research, Wako-shi, Saitama 351-0198, Japan
}

\begin{abstract}
We propose a quantum-classical hybrid algorithm to simulate the non-equilibrium steady state of an open quantum many-body system, named the dissipative-system Variational Quantum Eigensolver (dVQE).
To employ the variational optimization technique for a unitary quantum circuit, we map a mixed state into a pure state with a doubled number of qubits and design the unitary quantum circuit to fulfill the requirements for a density matrix.
This allows us to define a cost function that consists of the time evolution generator of the quantum master equation.
Evaluation of physical observables is, in turn, carried out by a quantum circuit with the original number of qubits. 
We demonstrate our dVQE scheme by both numerical simulation on a classical computer and actual quantum simulation that makes use of the device provided in Rigetti Quantum Cloud Service.
\end{abstract}
\date{\today}

\maketitle

\section{Introduction}\label{sec:introduction}
Technological developments in quantum devices have now reached a stage to realize the near-term quantum computers that contain from tens to hundreds of qubits with high gate fidelity, although their qubits are prone to error (or not fault-tolerant)~\cite{krantz_2019, bruzewicz_2019, aspuru_2012}.
Such quantum computers are dubbed as the Noisy Intermediate-Scale Quantum (NISQ) devices~\cite{preskill_2018}.
Even though their rigorous computational speedup is still obscured,
it is becoming practically impossible to simulate the NISQ devices by classical computers~\cite{harrow_2017, boixo_2018, chen_2018, bouland_2018, bravyi_2018, villalonga_2019, bravyi_2019}.
This triggers a surging interest in utilizing the NISQ devices for various kinds of problems in the real world:
the simulations of quantum many-body systems including their dynamics~\cite{georgescu_2014},  quantum chemistry calculations~\cite{mcardle_2018}, combinatorial optimization problems~\cite{farhi_2014}, and so on. 

In particular, intriguing research field among the proposed applications of the near-term quantum computers are the Variational Quantum Eigensolver (VQE) scheme~\cite{peruzzo_2014} and its variants.
Originally proposed for obtaining the ground state and its energy of a given Hamiltonian, the extensions of the VQE scheme are also capable of the excited states and their energies~\cite{macclean_2017QSE, nakanishi_2018, colless_2018, tyson_2019, higgott_2019, parrish_2019}.
So far, such developments including actual quantum simulations~\cite{aspuru_2005, omalley_2016, kandala_2017, hempel_2018} focus only on {\it closed} quantum systems,
which is a system that does not exchange energy with its external environment.
Physical systems in reality inevitably interact with their environment, and full understanding of such quantum phenomena is essential, e.g., for exploring the rich nature at out-of-equilibrium and designing quantum information processing devices.

A particularly significant topic is the non-equilibrium steady state (NESS) under time-independent dissipation~\cite{cai_2013, cui_2015, werner_2016, kshetrimayum_2017, yoshioka_2019, hartmann_2019, nagy_2019, vincentini_2019, weimer_2015, jin_2016, jin_2018, weimer_2019}.
For instance, it has been shown that 
properly engineered dissipations can stabilize  intriguing quantum states that are topologically nontrivial~\cite{diehl_2011} or useful for measurement-based quantum computation~\cite{kraus_2008, verstraete_2009}.
Also, the transport property of the NESS in nano-scale devices such as single-atom junctions is an important topic~\cite{dubi_2008}.
Such surging interests in the totally new regime due to the variety of behaviors are also reflected in the development of variational approaches for the NESS based on, e.g., tensor network techniques~\textcolor{black}{\cite{verstraete_2004, zwolak_2004, cai_2013, cui_2015, werner_2016, kshetrimayum_2017}} and neural networks~\cite{yoshioka_2019, hartmann_2019, nagy_2019, vincentini_2019}.
We note that several proposals on the NISQ devices consider open systems~\cite{yuan_2018,endo_2018, hu_2019},
but their focuses are on the dynamics and
no direct treatment of the NESS has been presented.

In this paper, we propose a scheme to calculate the NESS of a given open system by using the methodology of the VQE.
The proposed method, which we refer to as the dissipative-system VQE (dVQE), allows one to variationally determine the NESS of the system by minimizing a bounded cost function that is closely related with the accuracy of the optimization.
To satisfy the necessary conditions required for mixed quantum states, 
a sophisticated ansatz, which uses doubled number of qubits to map a mixed state into a pure state, is employed in the dVQE scheme.
Since a naive procedure to calculate the physical observables for the NESS becomes exponentially inefficient as the number of qubits increases, a measurement technique is introduced to avoid this problem.
We demonstrate our dVQE scheme by two styles of simulation; one is performed solely on a classical computer~\cite{qulacs_2018}, and another is executed in actual quantum-classical hybrid manner that makes use of the NISQ device provided in Rigetti Quantum Cloud Service~\cite{rigetti_2016}.

Both VQE and dVQE algorithms are variational algorithms that share the strategy to accomplish the speed-up: minimizing the expectation value of some Hermitian operator via optimization of variational quantum circuits. 
It is not so bold to expect that,  when NISQ devices become powerful enough to simulate classically intractable isolated systems via the VQE algorithm with sufficient fidelity,
stationary states that can be described with similar number of quantum gates can also be tackled using the dVQE algorithm.
\textcolor{black}{
Meanwhile, rigorous discussion on  condition to achieve speed-up over classical algorithms remains to be an open question. 
We also mention that the scaling of errors in quantum states and physical observables is a subtler issue in the dVQE algorithm compared to that in VQE, and therefore careful analysis by either numerics or experiments is desirable.
}

The remainder of the paper is organized as follows.
In Sec.~\ref{sec:openquantum}, we introduce the time independent quantum master equation for mixed states. In particular, the generator of the time evolution, or the Liouvillian, for the vectorized expression of the density matrix is discussed. We map the problem of searching the non-equilibrium steady state into an optimization problem in Sec.~\ref{sec:method}, and also discuss the appropriate variational quantum circuit and measurement scheme for the density matrix.
After we provide both numerical and experimental demonstrations of our dVQE scheme in Sec.~\ref{sec:simulation}, we present conclusion and discussion in Sec.~\ref{sec:conclusion}. \textcolor{black}{For completeness, we discuss the difference of distance measures in Appendix~\ref{sec:distance}.}
In Appendix~\ref{sec:seqmin}, we provide the outline of the sequential minimal optimization method \cite{nakanishi_2019}. 
The procedure to mitigate the gate error in the quantum circuit is provided in Appendix~\ref{sec:mitigation}, \textcolor{black}{and the numerical demonstration of the dVQE algorithm without any sampling noise or gate error is provided in Appendix~\ref{sec:exact_simulation}.}

\section{Equivalent representations of quantum master equation}\label{sec:openquantum}
Among the numerous schemes that treat the continuous-time open quantum systems, here we choose the standard formalism which is described by the homogeneous Markovian master equation. 
Such evolution of a mixed state $\rho$, which is assured to fulfill the completely-positive and trace-preserving property, is governed by the so-called Gorini-Kossakowski-Sudarshan-Lindblad (GKSL) equation \cite{lindblad_1976, gorini_1976} or equivalently the ordinary quantum master equation of the Lindblad type.
In the following, we give the expression of the GKSL equation in two equivalent ways: the matrix and vector representations. 
The former is the originally proposed quantum master equation for the density matrix, and the latter, initially adopted by Ref.~\cite{cui_2015} to obtain the NESS variationally, maps density matrix into a pure state with doubled number of qubits.

\subsection{Matrix representation}\label{subsec:matrix_representation}
First, we introduce the ordinary matrix representation of the GKSL equation.
The Liouvillian superoperator  $\mathcal{L}: \mathcal{H} \otimes \mathcal{H} \rightarrow \mathcal{H} \otimes \mathcal{H}$ for the finite-dimensional Hilbert space $\mathcal{H}$ governs the time evolution as follows,
\begin{eqnarray}\label{eqn:lindblad}
\dot{\rho} = {\mathcal L}\rho := - i[ H , \rho ] + \sum_k\frac{\gamma_k}{2} {\mathcal D}[c_k]\rho,
\end{eqnarray}
where the commutator in the right-hand side, $[A,B] = AB-BA$, describes unitary part of the dynamics given by the Hamiltonian of the system $H:\mathcal{H} \rightarrow \mathcal{H}$. 
The second term with dissipation amplitude $\gamma_k$ destroys the coherence of the pure states in general, governed by a superoperator $\mathcal{D}[c_k]:\mathcal{H}\otimes \mathcal{H} \rightarrow\mathcal{H} \otimes \mathcal{H}$ that acts on an operator as 
\begin{eqnarray}\label{eqn:superoperator}
{\mathcal D}[c_k]\rho &=&  c_k \rho c_k ^ { \dagger } - \frac { 1 } { 2 } \left\{c_k^{\dagger} c_k , \rho \right\},
\end{eqnarray}
where $c_k$ is the $k$-th jump operator which determines the detail of dissipation.
In the following, we consider time-independent Liouvillian superoperator which has at least one steady state that satisfies
\begin{eqnarray}
\mathcal{L}\rho_{\rm SS} = 0,
\end{eqnarray}
where $\rho_{\rm SS}$ denotes the density matrix of the NESS \cite{rivas_2011}.

\subsection{Vector representation}\label{subsec:vector_representation}
Although universal quantum computers may simulate completely-positve and trace-preserving maps so that the time evolution of quantum open systems can be studied~\textcolor{black}{\cite{kliesch_2011}}, this is not the case for the NISQ devices as same as simulation of Hamiltonian dynamics.
Therefore, we discuss the equivalent formalism for the GKSL equation in which  density ``matrix" $\rho:\mathcal{H}\rightarrow \mathcal{H}$ is mapped to a ``vector," namely a pure state in the Hilbert space with doubled number of qubits as $\ket{\rho} \in \mathcal{H}\otimes \mathcal{H}$.
\textcolor{black}{Concretely, we employ the Choi-Jamio\l kowski isomorphism which maps the quantum state as follows,}
\begin{eqnarray}\label{eqn:vector_map}
\rho = \sum_{ij} \rho_{ij} \ket{i}\bra{j}\ \ \mapsto \ket{\rho} = \sum_{ij}\frac{\rho_{ij}}{C}\ket{i}_{\mathcal{P}} \otimes \ket{j}_{\mathcal{A}},
\end{eqnarray}
where the factor $C = \sqrt{\sum_{ij}|\rho_{ij}|^2}$ assures the norm of the state to be unity.
Such a mapping is essential in application of the NISQ devices which is, in general, only capable of unitary operations on pure states.
Here, we discriminate the qubits that correspond to the ket and bra in the matrix representation by referring to them as the physical and ancillary qubits, respectively. \textcolor{black}{As in Eq.~\eqref{eqn:vector_map}, this is reflected on the subscripts $\mathcal{P}$ and $\mathcal{A}$.}
Although two representations of GKSL equation yield identical results, we emphasize that the normalization of states differs from each other; the trace of the matrix is set to unity in the matrix representation, i.e., $\sum_i \rho_{ii} = 1$, while the L$_2$ norm of $\ket{\rho}$ is unity in the vector representation. 

Under the mapping defined in Eq.~\eqref{eqn:vector_map}, the Liouvillian superoperator acts as linear operator on the Hilbert space with doubled number of qubits. 
Such generator in the vector representation $\hat{\Lind}:(\mathcal{H}\otimes \mathcal{H} \rightarrow \mathcal{H} \otimes \mathcal{H})$, dubbed as the Liouvillian operator in the following, acts on a state $\ket{\rho}$ as
\begin{eqnarray}
\hat{\mathcal{L}}\ket{\rho}
&=&
\left(-i(H\otimes \mathbbm{1} - \mathbbm{1}\otimes H^T) + \sum_k\hat{\mathcal{D}}[c_k]\right)\ket{\rho},  \\
\hat{\mathcal{D}}[c_k]& = & \frac{\gamma_k}{2}\left(c_k \otimes c_k^* - \frac{1}{2} c_k^{\dagger}c_k \otimes \mathbbm{1} - \mathbbm{1} \otimes \frac{1}{2} c_k^T c_k^* \right),
\label{eqn:liouvillian_vector}
\end{eqnarray}
where $*$ denotes the complex conjugate.
This can be derived by utilizing the mapping of operations that act from left and right of the density matrix as
\begin{eqnarray}\label{eqn:left_right_operator}
A \rho B \mapsto |A\rho B\rangle  =  A \otimes B^T \ket{\rho}.\label{eqn:A_rho_B_in_vector_representation}
\end{eqnarray}
Using the new representation, the problem of finding the NESS is expressed in terms of standard linear algebra. Concretely, our goal is to solve the following equation for non-hermitian operator $\hat{\mathcal{L}}$ given as
\begin{eqnarray}\label{eqn:NESS_vector}
\hat{\mathcal{L}}|\rho_{\rm SS}\rangle = 0,
\end{eqnarray}
\textcolor{black}{or equivalently }
\begin{eqnarray}\label{eqn:NESS_vector}
\textcolor{black}{\hat{\mathcal{L}}^{\dagger}\hat{\mathcal{L}}|\rho_{\rm SS}\rangle = 0.}
\end{eqnarray}
\textcolor{black}{As we discuss in detail in Sec.~\ref{sec:method}, the expectation value of $\hat{\mathcal{L}}^{\dagger}\hat{\mathcal{L}}$ can be considered to be a proper choice of the cost function to compute the NESS in NISQ devices.}

Two remarks are in order before proceeding to the next section. 
Firstly, the vector representation is different from the purification since one do not retrieve the original mixed state by tracing out the ancillary qubits.
Rather, each matrix elements are individually embedded as the amplitude of $\ket{\rho}$. 
This is crucial for reformulating the Liouvillian as a state-independent operator, and therefore for casting the problem into a variational optimization problem for NISQ devices.
Secondly, we may also formalize the problem using the purification of the mixed state. 
In such a case, we search for the state which is invariant under the non-unitary time evolution generated by the Liouvillian superoperator.
However, it is non-trivial whether such time evolution operation could be expressed by polynomial number of Pauli operators.

\section{Method}\label{sec:method}
In the following, we introduce the dVQE scheme to simulate the NESS and its physical observables in open quantum systems. 
This is done by firstly casting the problem of finding the NESS into an optimization problem of an appropriate Hermitian operator in the vector representation. 
Next, the structure of the variational quantum circuit suited to express the density matrix in the vector representation is provided, and finally the measurement procedure for  the physical observables in the original matrix representation is discussed. 
Such modification of the representation is crucial to execute proper calculation.

\subsection{Searching NESS by variational optimization}
\label{subsec:variationalNESS}
The Liouvillian operator defined as Eq.~\eqref{eqn:liouvillian_vector} is in general a non-hermitian matrix.
If the dissipation is absent or parity-time symmetric~\cite{bender_2007}, the Liouvillian operator is skew hermitian, and therefore is diagonalizable by an unitary matrix to obtain pure imaginary eigenvalues. 
\textcolor{black}{A generic Liouvillian may not be diagonalizable at all; the Jordan normal form might consist of non-trivial Jordan blocks. The eigenvalues are complex in general, and the eigenvectors, if one may compute, are not orthogonal.}

In contrast, the product with the adjoint of the Liouvillian operator, $\hat{\mathcal{L}}^{\dagger} \hat{\mathcal{L}}$, is a Hermitian matrix with non-negative spectrum. 
\textcolor{black}{It is straightforward to see that the NESSs, the kernel of $\hat{\mathcal{L}}$, correspond to the eigenstate(s) with lowest eigenvalue(s) $\lambda = 0$. More explicitly, the NESS $\ket{\rho_{\rm SS}} $ satisfies}
\begin{eqnarray}\label{eqn:NESS_vector_withconj}
\hat{\mathcal{L}}^{\dagger} \hat{\mathcal{L}} \ket{\rho_{\rm SS}} = 0.
\end{eqnarray}
In the following, we exclusively consider models that can be shown to have unique NESS~\cite{schirmer_2010} so that the ground-state-search techniques for the variational quantum circuit can be applied.
In the dVQE scheme, we employ the expectation value of $\hat{\mathcal{L}}^{\dagger} \hat{\mathcal{L}}$ as the cost function. The 
parameters in the ansatz are optimized to reach the minimal value of the cost function which is lower-bounded by zero.

The task of optimization is divided into two processing units; the QPU (quantum processing unit) carries the quantum state while the CPU executes the optimization of the variational parameter based on the measurement result in the QPU~\cite{peruzzo_2014}. 
A quantum-classical hybrid approach designed for open quantum systems can be summarized as follows.
\begin{enumerate}
    \item[1.] Initialize the variational parameters of the ansatz $U(\theta)$.
    \item[2.] QPU part: Estimate the cost function $\braket{\hat{\mathcal{L}}^{\dagger} \hat{\mathcal{L}}}$
    by sampling from the quantum state given by the variational quantum circuit as $\ket{\rho_{\theta}} = U(\theta)\ket{0}^{\otimes 2N}$.
    \item[3.] CPU part: Based on the measurement results in QPU, compute the new parameters $\theta'$ using a classical optimization technique and replace $\theta$.
    \item[4.] Repeat 2 and 3 until the cost function converges.
\end{enumerate}

Before proceeding to discuss the variational quantum circuit, we provide three remarks in order.
First, the dVQE scheme can be straightforwardly extended to models with multiple NESSs by using the VQE methods that deal with excited states~\cite{macclean_2017QSE, nakanishi_2018, colless_2018, tyson_2019, higgott_2019, parrish_2019}.
In general, the largest or smallest singular values of an operator given by a sum of local operator can be obtained by the proposed method.
Secondly, while the lower bounded property of the cost function, $\hat{\mathcal{L}}^{\dagger} \hat{\mathcal{L}}$, gives us information on the quality of the optimization~\footnote{It can be shown that the overlap in the vector representation, $f$, between the exact NESS and the optimized ansatz  can be bounded as $1 - f^2\geq \braket{\mathcal{L}^{\dagger} \mathcal{L}}/\delta$ where $\delta$ is the lowest non-zero eigenvalue of $\hat{\mathcal{L}}^{\dagger}\hat{\mathcal{L}}$.
}, one must be aware of the difference of the normalization between the matrix and vector representation
 when one sets the desirable value of the cost function.
\textcolor{black}{If the optimization in the vector representation is done accurate enough so that $\langle \braket{\rho_{\theta}| \hat{\mathcal{L}}^{\dagger}\hat{\mathcal{L}}|\rho_{\theta} }\rangle < \epsilon$ is satisfied, then the error for any physical observable $A$ can be given as $|{\rm Tr}[\rho_{\theta} A] - {\rm Tr}[\rho_{\rm SS}A]| < f(\epsilon)$. Here, $f(\epsilon)$ clearly becomes zero if $\epsilon = 0$, although the scaling of $f$ cannot be determined straightforwardly. 
See Appendix \ref{sec:distance} for discussion based on numerical calculation.
We leave mathematical proof as an important open problem.}
Thirdly, weighing the number of the sampling shots in accordance with the amplitude of each coefficient is expected to play an important role to reduce the total number of measurement.
One also may consider using the grouping of simultaneously measurable Pauli strings~\cite{jena_2019, yen_2019, izmaylov_2019, izmaylov_2019_revising, rubin_2018, verteletskyi_2019}, while its advantage is not assured in general. 
If {\it both} the coefficient and expectation value of each Pauli term are taken into account, then the grouping strategy is guaranteed to be beneficial~\cite{crawford_2019}.

\subsection{Ansatz for dVQE}\label{subsec:ansatz}
In the following, we introduce a variational quantum circuit to represent the NESS in the vector representation, which, as we have seen in \S\ref{subsec:variationalNESS}, enables us to execute the optimization in a parallel fashion with the VQE for closed systems \cite{peruzzo_2014}.
Note, however, that the following conditions are required for the density matrix $\rho$ to be physical:
\begin{enumerate}
    \item[(I)] $\rho^{\dagger} = \rho$ (Hermiticity),
    \item[(II)] $\braket{\psi|\rho|\psi} \geq 0, ^\forall \ket{\psi}\in\mathcal{H}$ (Positive semi-definiteness),
    \item[(III)] ${\rm Tr}[\rho] = 1$ (Unit trace).
\end{enumerate}
As we describe in the following, our dVQE scheme satisfies all the conditions. 
It suffices to consider the condition (III) only when we measure the expectation value of physical observables; the normalization of the representation can be neglected during the optimization without affecting the accuracy of the optimization.
In the following, we discuss how to design an appropriate ansatz to satisfy the conditions (I) and (II)~\textcolor{black}{\footnote{Note that it is a NP-hard problem to check whether the condition (II) is satisfied in tensor network methods~\cite{kliesch_2014}.}}.

\begin{figure}[t]
\begin{center}
\begin{tabular}{c}
  \begin{minipage}{0.97\hsize}
    \begin{center}
     \resizebox{0.95\hsize}{!}{\includegraphics{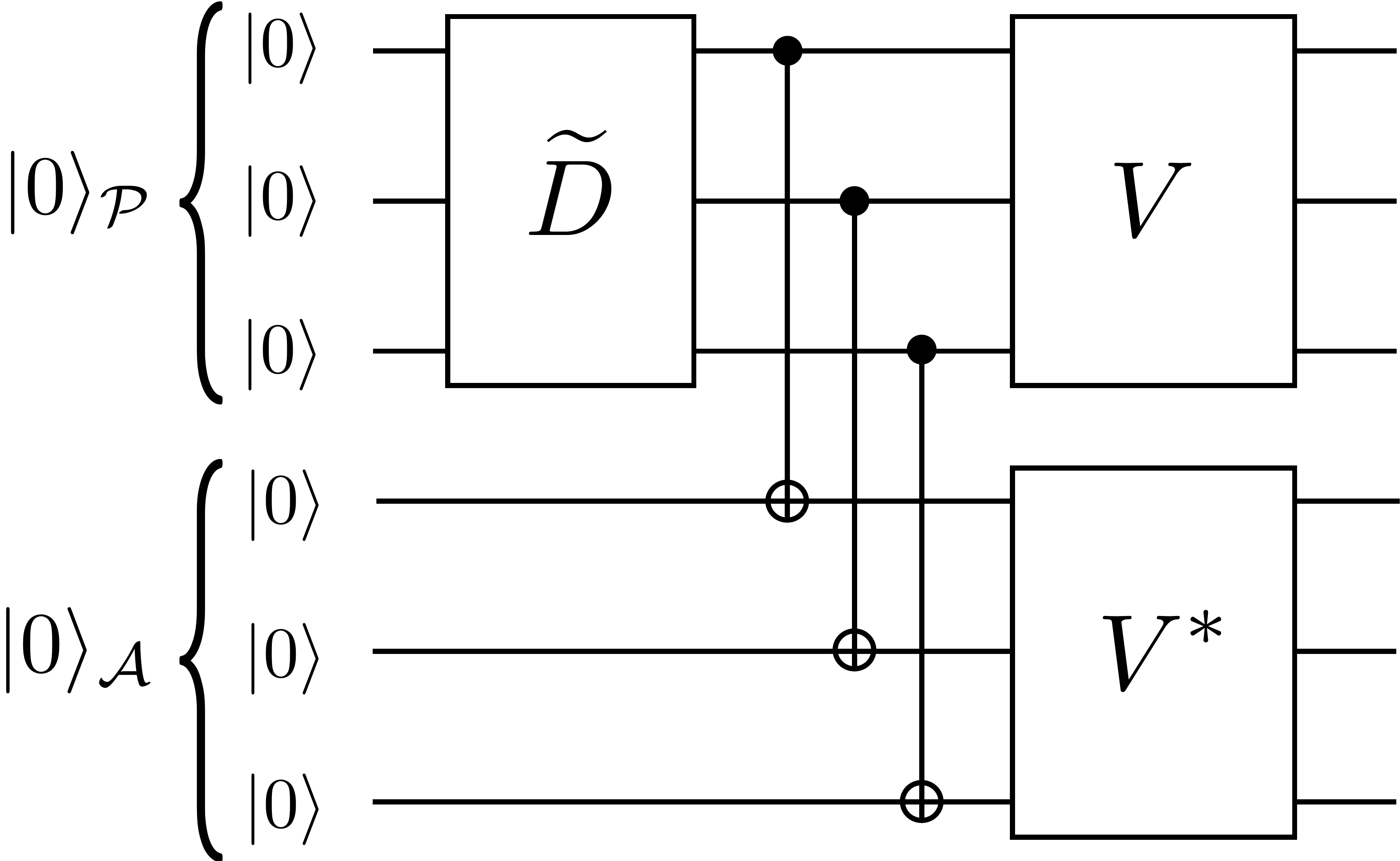}}
   \end{center}
  \end{minipage}
 
\end{tabular}
\end{center}
\caption{\label{fig:ansatz_general} (Color Online) Variational quantum circuit as the ansatz for the vector representation of the NESS of open quantum system with 3 physical qubits. Both the physical and ancillary qubits, denoted as $\ket{0}_{\mathcal{P}}$ and $\ket{0}_{\mathcal{A}}$ respectively, are initialized to zero. To impose the Hermiticity and positive-semidefiniteness that are required for density matrices, quantum gates for eigenvalue distribution $\widetilde{D}$, CNOT gates between physical and ancillary qubits, and basis transformation $V$ operate sequentially. 
}
\end{figure}

We first diagonalize the density matrix $\rho$
using unitary matrix.
Let $q_k$ be the $k$-th bit that takes either 0 or 1 and ${\bf q} = (q_1, \cdots, q_{N})$ be an array of $N$ bits, and we decompose the density matrix as
\begin{eqnarray}
\rho &=& V D V^{\dagger}, \label{eqn:dm_decomposition}\\
D &=& {\rm diag}\left(\{\lambda_{\bf q} \} \right).
\end{eqnarray}
Here, $D$ 
gives the eigenvalue distribution, which are related to the entropy of the state, and $V$
\textcolor{black}{is a generic unitary transformation that transforms the basis into a desired one.} 
As is graphically described in Fig.~\ref{fig:ansatz_general}, the vector representation of Eq.~\eqref{eqn:dm_decomposition} is given by using the mapping given by Eq.~\eqref{eqn:A_rho_B_in_vector_representation} as
\begin{eqnarray}
\ket{\rho} &=& \left[V\otimes V^*\right] \ket{D}
\end{eqnarray}
where the eigenvalue distribution is expressed as
\begin{eqnarray}
\ket{D} &=& \sum_{\bf q} \frac{\lambda_{\bf q}}{C} \ket{\bf q}_{\mathcal{P}} \otimes \ket{\bf q}_{\mathcal{A}} \nonumber \\
&=& \left(\prod_{n=1}^{N}{\mathrm{CNOT}}_{n, n+N} \right) 
\sum_{\bf q}\frac{\lambda_{\bf q}}{C} \ket{\bf q}_{\mathcal{P}} \otimes \ket{0}_{\mathcal{A}},\\
&=& \left(\prod_{n=1}^{N}{\mathrm{CNOT}}_{n, n+N} \right)\widetilde{D}|0\rangle_{\mathcal{P}}\otimes |0\rangle_{\mathcal{A}}.
\end{eqnarray}
Here, CNOT gates operating on the $n$-th and $(n+N)$-th qubits, which are taken as the control and target qubits, respectively, entangle the left and right space such that the matrix representation of the state yields a diagonal matrix. 
Viewed in the computational basis, CNOT gate is concretely expressed as
\begin{eqnarray}
{\rm CNOT} = \ket{0}\bra{0}\otimes\left(\ket{0}\bra{0} + \ket{1}\bra{1}\right) +\nonumber \\ \ket{1}\bra{1}\otimes\left(\ket{0}\bra{1} + \ket{1}\bra{0}\right),
\end{eqnarray}
where the first and second qubits denote the control and target qubits.

We propose an ansatz based on Eq.~\eqref{eqn:dm_decomposition} as 
\begin{eqnarray}
\ket{\rho_{\theta}} &=& U(\theta)\ket{0} \\
&=&\left[V(\theta_v) \otimes V^*(\theta_v)\right] \nonumber \\
&&\ \ \ \times\left(\prod_{n=1}^{N}{\mathrm{CNOT}}_{n, n+N}\right)
\widetilde{D}(\theta_d) \ket{0},\label{eqn:ansatz_total}
\end{eqnarray}
where $U(\theta)$ is the ansatz for the NESS that consists of  quantum circuit $\widetilde{D}(\theta_d)$ ($V(\theta_v)$) for the eigenvalue distribution (basis transformation) parametrized by a set of $R_d$ ($R_v$) values, i.e., $\theta_d := \{\theta_d^r\}_{r=1}^{R_d}$ ($\theta_v:=\{\theta_v^r\}_{r=1}^{R_v}$).
Here, the union of the variational parameters are denoted as $\theta = \theta_d \cup \theta_v$. 
The matrix representation of a state given by Eq.~\eqref{eqn:ansatz_total} automatically fulfills the condition (I), and also the condition (II) can be satisfied by imposing appropriate restriction on the parameters $\theta_v$.

\begin{figure}[t]
\begin{center}
\begin{tabular}{c}
  \begin{minipage}{0.97\hsize}
    \begin{center}
     \resizebox{0.95\hsize}{!}{\includegraphics{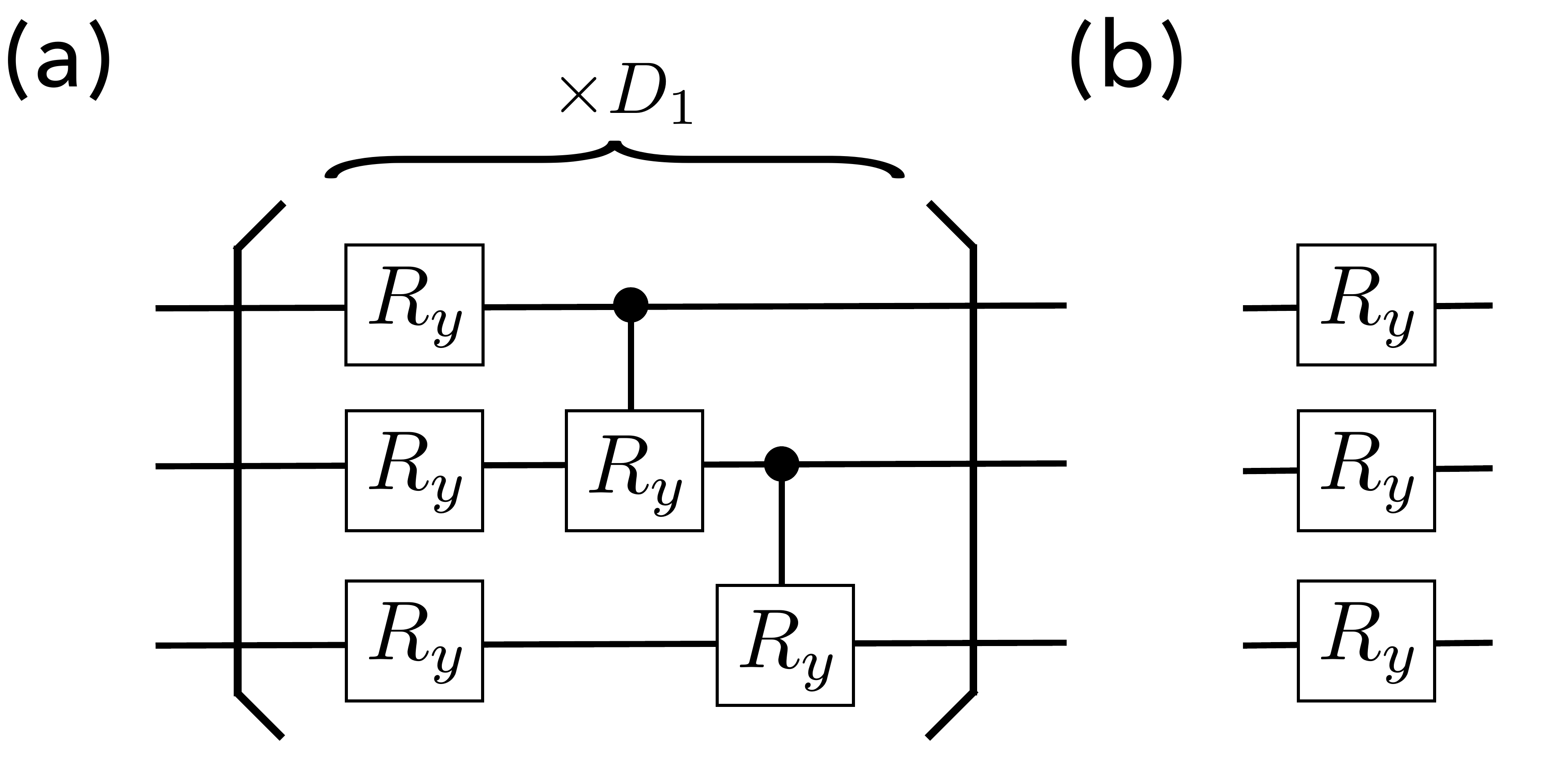}}
   \end{center}
  \end{minipage}
 
\end{tabular}
\end{center}
\caption{\label{fig:eigval_ansatz} (Color Online) Examples for the quantum circuit for eigenvalue distribution $\widetilde{D}$. (a) The entangled type which consists of the repeating structure of single-qubit gates and controlled Y rotation. Here, $D_1$ denotes the repetition of the block in the bracket to enhance the representability of the ansatz. (b) The decoupled type with solely the single-qubit Y rotations. 
}
\end{figure}

Shown in Fig.~\ref{fig:eigval_ansatz} are two concrete examples for $\widetilde{D}(\theta_d)$: one with a repeating structure of single-qubit and controlled Y rotations and another with solely the single-qubit Y rotation. Such quantum circuits, graphically shown in Fig.~\ref{fig:eigval_ansatz} (a) and (b), respectively, are referred to as the ``entangled type" and ``decoupled type" in the following. 
Here, the unitary gate for the rotation along $a$-axis is given as $e^{-i\sigma_a\theta/2}$ where $\theta$ denotes the \textcolor{black}{rotation} angle.
\textcolor{black}{While the entangled ansatz becomes increasingly more powerful along the depth, the analytical expression with respect to circuit parameters becomes intractable. 
As we mention in Sec.~\ref{subsec:measurement}, this requires us to estimate the eigenvalue distribution $\{\lambda_{\bf q}\}$ when physical observables are to be computed. The decoupled ansatz, on the other hand, provides the exact expression of each $\lambda_{\bf q}$. While this is expected to be beneficial in improving the accuracy of the calculation, the number of parameters is limited and hence may not be suited for states with extensive entropy.}
\textcolor{black}{One may also employ multi-qubit gates with positive real values to increase the expressive power of the ansatz.}

\begin{figure}[t]
\begin{center}
\begin{tabular}{c}
  \begin{minipage}{0.97\hsize}
    \begin{center}
     \resizebox{0.95\hsize}{!}{\includegraphics{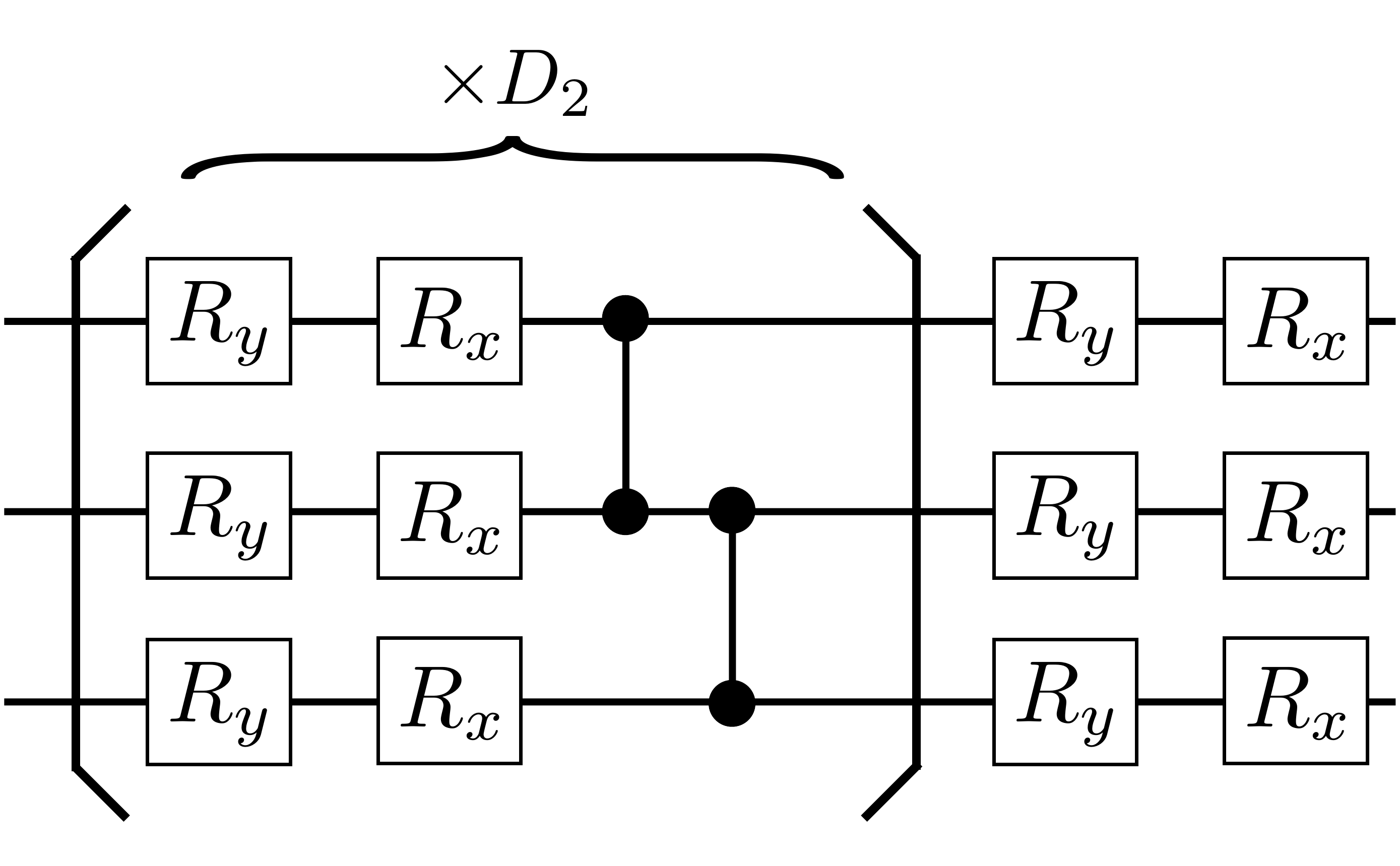}}
   \end{center}
  \end{minipage}
 
\end{tabular}
\end{center}
\caption{\label{fig:basis_ansatz} (Color Online) Example for the quantum circuit for the basis transformation $V$. Here, we adopt the so-called hardware-efficient ansatz which repeats the block consisting of an arbitrary single-qubit gate for each qubit and CZ gates between neighboring qubits. The repetition number of the block in the bracket is denoted as  $D_2$.
}
\end{figure}

For the basis transformation $V(\theta_v)$, we adopt the so-called hardware-efficient ansatz~\cite{kandala_2017} as we show in Fig.~\ref{fig:basis_ansatz}. The block consisting of an arbitrary single-qubit gate for each qubit and CZ gates between neighboring qubits are repeated.

\subsection{Measurement of physical observables}\label{subsec:measurement}
After the optimization scheme proposed in Sec.~\ref{subsec:variationalNESS} has been executed for the variational quantum circuit introduced in Sec.~\ref{subsec:ansatz}, our interest would be focused on the measurement of physical observable for the NESS.

Note that the vector representation is not appropriate when evaluating observables. 
Any measurement of an observable defined on a state $\ket{\rho} = \sum_{ij}\frac{\rho_{ij}}{C}\ket{i}_{\mathcal{P}}\ket{j}_{\mathcal{A}}$ that represents a density matrix $\rho$ would involve products of the coefficients $\rho_{ij}$, which is incompatible to, for example, the expression of an expectation value of a physical observable $O$ computed as  $\mathrm{Tr}(\rho O)$.

To avoid the above-mentioned problem, we map the optimized ansatz into the matrix representation which requires only $N$ qubits instead of $2N$ qubits. 
With the dependence on the variational parameters abbreviated for simplicity,  the optimized ansatz is mapped from
$\ket{\rho} = \left[V\otimes V^* \right] \widetilde{D} \ket{0}$ into $\rho = \sum_{\bf q}\lambda_{\bf q} V\ket{\bf q}\bra{\bf q} V^{\dagger}.$
As is the case for the experiments shown in the next section, the summation over all computational basis can be done exactly when the number of qubits $N$ is sufficiently small; \textcolor{black}{we may numerically compute $\lambda_{\bf q}$ from the parameters of $\widetilde{D}$ and weigh $\bra{\bf q}V^{\dagger}O V\ket{\bf q}$ accordingly to obtain}
\begin{eqnarray}
\braket{O} &=& {\rm Tr}[\rho O]\\
&=& \sum_{\bf q} \lambda_{\bf q} \braket{{\bf q}| V^{\dagger} O V|{\bf q}}.\label{eqn:obs_cal}
\end{eqnarray}

When $N$ becomes larger, taking the summation over $\bf q$'s becomes intractable and hence is replaced by stochastic sampling. 
\textcolor{black}{Let us assume a variational circuit with the decoupled type ansatz as in Fig.~\ref{fig:eigval_ansatz}(b) for the eigenvalue distribution and hard-ware efficient ansatz for basis transformation. 
In such a case, we can generate a state $\ket{\bf q}$ with likelihood $\lambda_{\bf q}$ 
by simply dephasing the quantum state $\widetilde{D}'\ket{0}$ where $\widetilde{D}'$ denotes the quantum gate with its variational angles $\theta_d$ replaced by $\theta'_d = \arccos\left(\sqrt{1/(1+\tan\theta_d)}\right)$.}
The observables are computed by repetitively carrying out the measurement for $V\ket{\bf q}$ until convergence.
 \textcolor{black}{In general,} we first estimate $\{\lambda_{\bf q}\}$ via sampling, and then measure the observable where the initial state is set in accordance with the probability distribution. 
As in ordinary VQE simulations, the number of measurements scales as $\mathcal{O}(1/\epsilon^2)$ given a target precision $\epsilon$ of the observable.

Finally, we summarize the measurement scheme using $N$ qubits as follows,
\begin{itemize}
\item[(0.] Optimize the set of variational parameters in the vector representation, i.e., using $2N$ qubits.)
\item[1.] Let $\bf q$ be an array of $N$ bits with nonzero probability $\lambda_{\bf q}$ and initialize the state of $N$ qubits to $\ket{\bf q}$.
\item[2.] Measure the observable $O$ using the quantum state with transformed basis $\ket{\psi_{\bf q}} = V\ket{\bf q}$.
\item[3.] Repeat 1 and 2 until the weighed sum $\braket{O} = \sum_{\bf q}\lambda_{\bf q}\braket{\psi_{\bf q}| O |\psi_{\bf q}}$ converges.
\end{itemize}
where $\lambda_{\bf q}$ is replaced with the approximated distribution if necessary.

\section{Results}\label{sec:simulation}
In the following, we give demonstration of our dVQE scheme by both quantum and numerical simulations of the quantum circuit with depolarizing error.
The cost function $\braket{\hat{\Lind}^{\dagger} \hat{\Lind}}$ is evaluated via sampling each Pauli term from noisy quantum circuit,
whose variational parameters are optimized by the sequential minimal optimization technique \cite{nakanishi_2019}. 
The detailed procedure is provided in Appendix~\ref{sec:seqmin}.
The mitigation scheme for the gate errors is also discussed in Appendix~\ref{sec:mitigation}.

As examples,  the quantum Ising model and the persistent current model~\cite{keck_2018} are simulated.
In both cases, we consider local dissipation such that the vector representation of the GKSL equation is given as
\begin{eqnarray}
\hat{\mathcal{L}}\ket{\rho} &=& 
\Bigg(-i(H\otimes \mathbbm{1} - \mathbbm{1}\otimes H^T)   \nonumber \\
&& \ \ \ \ \ \ \ + \sum_{a} \sum_{i=0}^{N-1} \gamma_a \hat{\mathcal{D}}[c^{(a)}_i]\Bigg)\ket{\rho}, \label{eqn:lindblad_ham}
\end{eqnarray}
where the $c_i^{(a)}$ denotes the $a$-th jump operator, with its amplitude $\gamma_a$, acting on the $i$-th site.

\subsection{Quantum Ising model}\label{subsec:1dTFIM}
\begin{figure}[t]
\begin{center}
\begin{tabular}{c}
  \begin{minipage}{0.97\hsize}
    \begin{center}
     \resizebox{0.95\hsize}{!}{\includegraphics{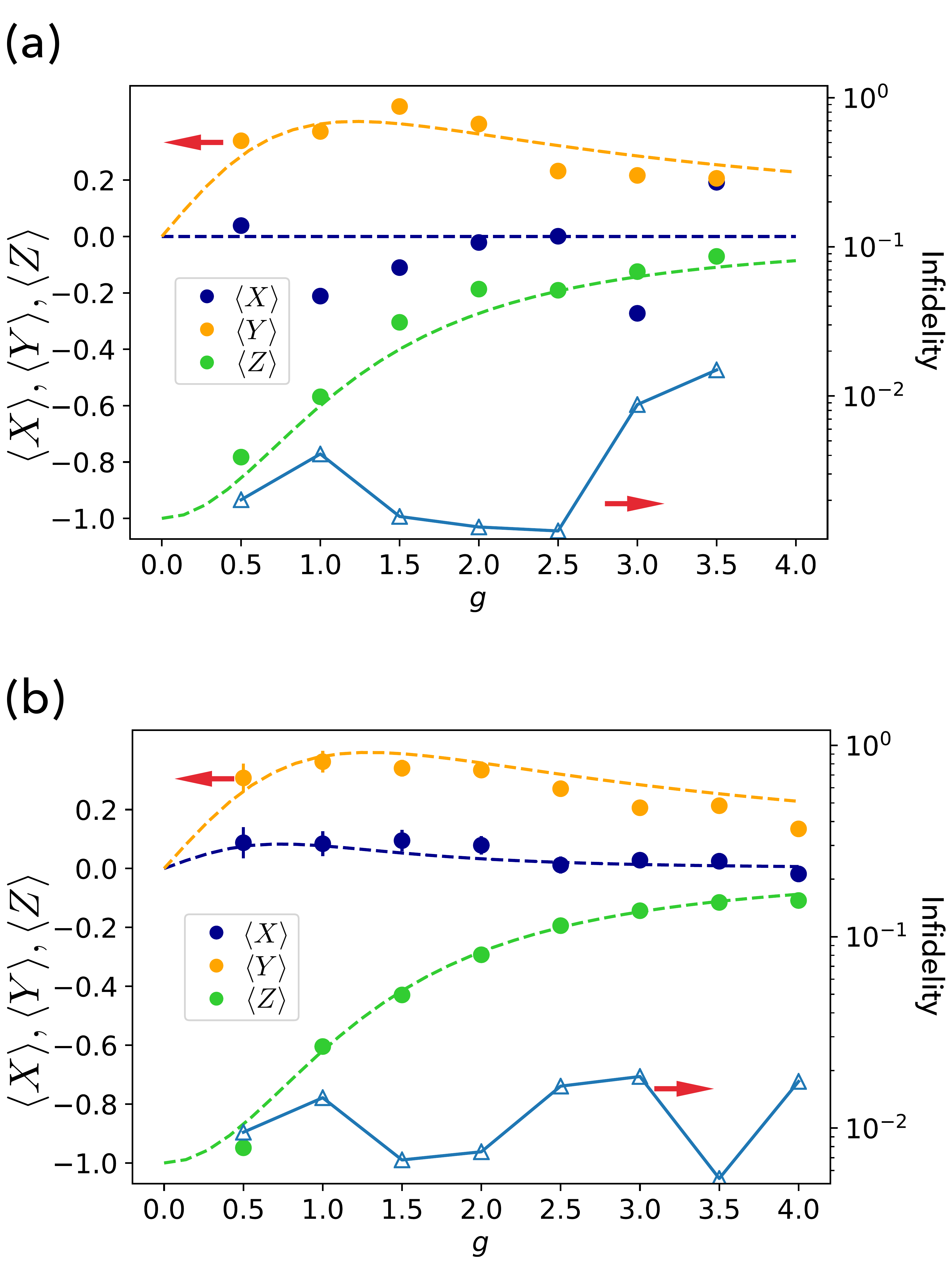}}
   \end{center}
  \end{minipage}
 
\end{tabular}
\end{center}
\caption{\label{fig:observable_1dTFIM_mitigated} (Color Online) (a) Quantum simulation and (b) \textcolor{black}{noisy} numerical simulation of magnetization curves (left axis) for the dissipative quantum Ising model.
The colors, blue, orange, and green, correspond to x,y, and z components of the mean magnetizations respectively for each dVQE (filled circles) and exact diagonalization (dotted lines).
 The unfilled triangles denote the infidelity $1 - F$ (right axis) which is typically $10^{-2}$. 
 \textcolor{black}{The red arrows indicate the corresponding axes for data points.}
 The dissipation amplitudes are taken as $\gamma_1 = 1$ and $\gamma_2 = 0.5$. \textcolor{black}{The eigenvalue distribution ansatz is decoupled (entangled) types shown pictorially in Fig.~\ref{fig:eigval_ansatz}(b) (Fig.~\ref{fig:eigval_ansatz}(a)) with the repetition numbers taken as $D_2 = 0$ ($D_1=D_2 = 1$) for system size $N=1(2)$ under quantum (numerical) simulation.} Each Pauli term is sampled over 400 times.
}
\end{figure}

First, we compute the NESS of the quantum Ising model with transverse field under both damping and dephasing effect.
The Hamiltonian for this model is defined as
\begin{eqnarray}
H &=& \frac{1}{2}\sum_{i} \sigma^z_i \sigma^z_{i+1} + g\sum_i \sigma^x_i. \label{eqn:liouvillian_1dTFIM}
\end{eqnarray}
 Here, $\sigma^{\alpha}_i$ $(\alpha=x,y,z)$ is the Pauli operator for the $i$-th spin and $g$ is the amplitude of the transverse field. 
The jump operators, denoted as $c^{(1)}_i$ and $c^{(2)}_i$ for the damping and dephasing effect on the $i$-th spin respectively, are given as
\begin{eqnarray}
c_i^{(1)} &=& \sigma^-_i, \\ 
c_i^{(2)} &=& \sigma^z_i,
\end{eqnarray}
with their strengths denoted as $\gamma^{(1)}$ and $\gamma^{(2)}$, respectively. The spin ladder operator is defined as $\sigma^{\pm}_i = (\sigma^x_i \pm i \sigma^y_i)/2$.
Although the model does not show any phase transition, the competition between the Ising interaction, transverse field and dissipation gives us suitable testbed for the dVQE scheme.

\textcolor{black}{Figure~\ref{fig:observable_1dTFIM_mitigated} shows the magnetization curves of the NESS with the dissipation amplitudes taken as $\gamma_1 = 1, \gamma_2 = 0.5$. 
The quantum (numerical) simulation is carried out for $N=1$ (2) site(s) for the decoupled (entangled) ansatz with repetition numbers taken as $D_2 = 0$ ($D_1=D_2 = 1$) such that there are 3 (11) parameters in total.
Note that in either simulation the quantities are evaluated via sampling under the presence of the noise, and hence the gate error mitigation has been performed for both the optimization and observable measurement. In Appendix ~\ref{sec:exact_simulation} we provide another numerical demonstration that excludes any gate noise or error mitigation to show its validity for $N=8$.}
From the density matrix obtained by the exact diagonalization and the dVQE scheme, the infidelity is calculated as $1- F$  where the fidelity $F$ between the two matrices, denoted as $\rho_1$ and $\rho_2$, respectively, is given as
\begin{eqnarray}
F = \left(\mathrm{Tr} \sqrt{\sqrt{\rho_{\rm ED}} \rho_{\rm dVQE} \sqrt{\rho_{\rm ED}}}\right)^2.
\end{eqnarray}
While the infidelity, which qualifies the accuracy of the optimized ansatz, is typically in the order of $10^{-2}$, the deviation of the observable is relatively larger. 
In particular, for the quantum simulation, we observe that the sequential minimization optimization applied in this paper can find optimal parameters quite accurately irrespective of the imperfect evaluation of the cost function.

\subsection{Coupled QED cavities with persistent current}\label{subsec:current}
Another good candidate for the demonstration of the dVQE scheme is the effective model of coupled QED cavities in the limit of  very strong repulsion, whose Hamiltonian is given as
\begin{eqnarray}
\textcolor{black}{H = \mu \sum_i \sigma_i^+ \sigma_i^-,}
\end{eqnarray}
where $\mu$ is the chemical potential which originates in the energy offset by the local repulsion of the photons.
In the following, we take the chemical potential as the scale of energy, i.e., $\mu = 1$.
It is proposed in Ref.~\cite{keck_2018} that, by appropriately engineering the interaction between the reservoir and the cavities, one may induce two-site dissipations that give rise to non-equilibrium current. 
Together with the damping and dephasing, we take such decoherence into account as
\begin{eqnarray}
c_i^{(1)} &=& \sigma^-_i, \nonumber\\ 
c_i^{(2)} &=& \sigma^z_i, \nonumber\\
c_i^{(3)} &=& \alpha \sigma_i^- + \beta \sigma_i^+ 
+ \gamma \sigma_{i+1}^- + \delta \sigma_{i+1}^+.\label{eqn:persistent_curr_jump}
\end{eqnarray}
The engineered dissipation, denoted as $c_i^{(3)}$, consists of four terms, and hence we set $\gamma^{(3)}=1$ for simplicity. 
To reduce the number of parameters, we set the parameters as $\alpha = \gamma^* = \cos\theta$ and $\delta = \beta^* = \sin\theta$.
In the following, we investigate the circulating current,
\begin{eqnarray}
\mathcal{I}^{\eta}_i = -\eta \sigma_i^+\sigma_{i+1}^- + {\rm h.c.},
\end{eqnarray}
where $\eta = |\alpha|^2 - |\delta|^2 = \cos^2\theta - \sin^2\theta$. 
This reservoir-engineering induced current does not vanish in the thermodynamic limit in contrast to the gauge-flux induced currents and also expected to be robust in the presence of perturbations. 

\begin{figure}[t]
\begin{center}
\begin{tabular}{c}
  \begin{minipage}{0.97\hsize}
    \begin{center}
     \resizebox{0.95\hsize}{!}{\includegraphics{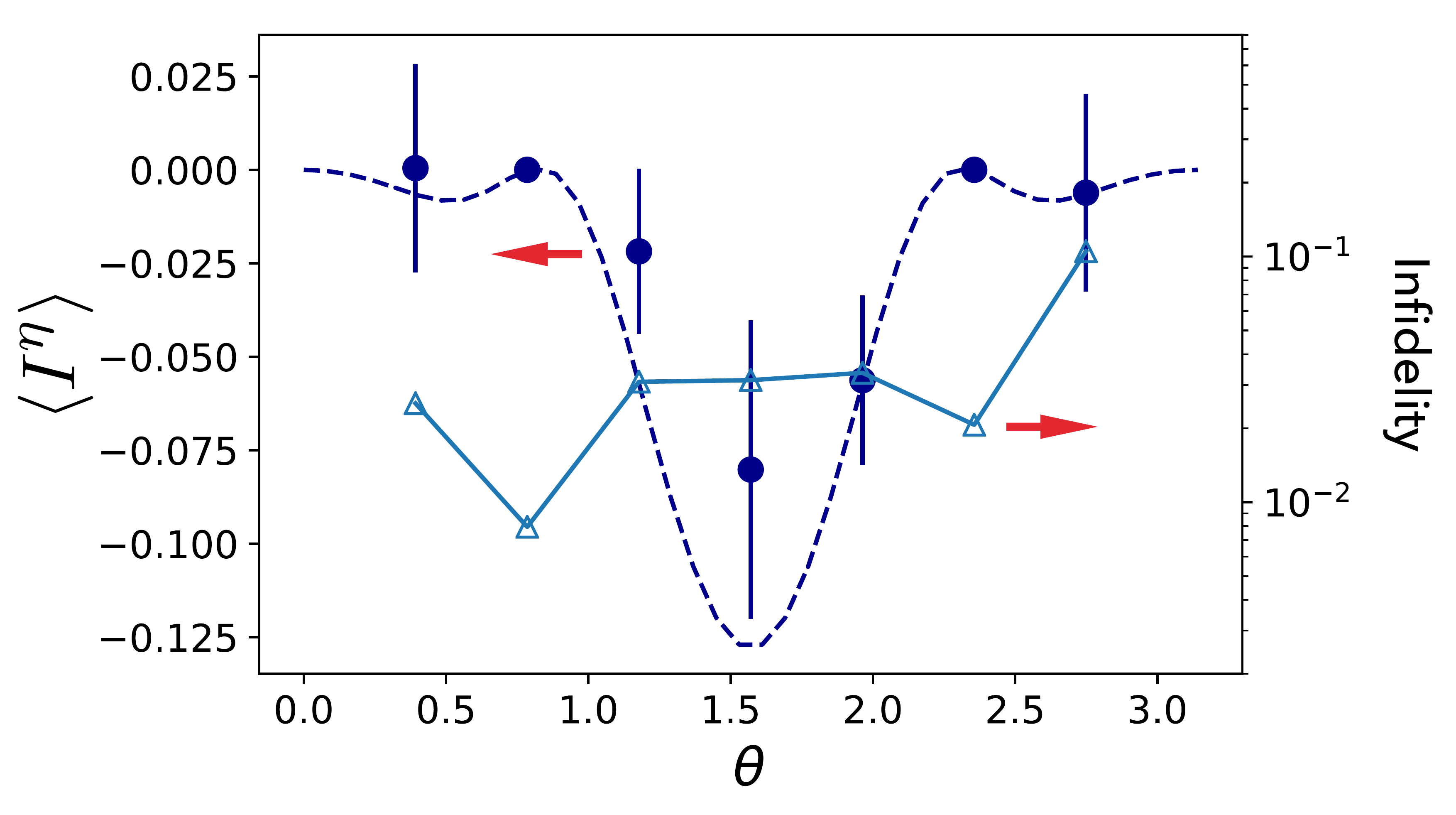}}

   \end{center}
  \end{minipage}
 
\end{tabular}
\end{center}
\caption{\label{fig:observable_cQED_N2} (Color Online)  Numerical simulation of the persistent current (left axis) and infidelity (right axis) for the reservoir-engineered coupled QED model. The blue filled circles and dashed line are the persistent current calculated by the dVQE scheme and exact diagonalization, respectively, and the unfilled triangles denote the infidelity. \textcolor{black}{The red arrows indicate the corresponding axes for the data points.}
The dissipation amplitudes are taken as $\gamma_1 = 0.3$ and $\gamma_2 = 0.5$ for damping and dephasing, respectively. We employ the entangled type for the eigenvalue distribution ansatz, and take the repetition numbers as $D_1 = D_2 = 2.$ Each Pauli term is sampled over 400 times.}
\end{figure}

\textcolor{black}{The result of numerical simulation for $N=2$ sites is shown in  Fig.~\ref{fig:observable_cQED_N2}. Here, the quantum circuit employs the entangled ansatz for the eigenvalue distribution and the circuit depth is taken as $D_1 = D_2 = 2$ such that there are 18 parameters in total.}
The persistent current creates an excitation on one site while destroying at the neighboring site, accurate evaluation of quantum entanglement is required compared to the case in the dissipative quantum Ising model.
\textcolor{black}{We observe that the magnitude of the current, reflecting that of two-body correlation, is relatively small and hence very sensitive to the noise. This causes the deviation of $\langle \mathcal{I}^{\eta}\rangle$ although the infidelity of the state is of $\order (10^{-2})$. }

\section{Conclusion}\label{sec:conclusion}
We have proposed the variational quantum algorithm to obtain the non-equilibrium steady state (NESS) of an open quantum many-body system, named the dissipative-system Variational Quantum Eigensolver (dVQE). Our dVQE scheme, based on the quantum-classical hybrid algorithm that are intended to compute the eigenstates and their eigenenergies of a given Hamiltonian, optimizes the variational parameters of the ansatz for the NESS by minimizing the cost function defined from the time evolution generator of the quantum master equation. Also a measurement scheme for the physical observables that characterize the quantum state has been included. 

The demonstration of the dVQE scheme is given by both quantum and numerical simulations. 
In the former, we have used the actual quantum device provided in Rigetti Quantum Cloud Service, and in the latter we have performed numerical simulations for noisy quantum circuits.
It is shown that the magnetization in dissipative quantum Ising model and non-equilibrium current in reservoir-engineered coupled QED cavity model can be simulated by the dVQE scheme.

We believe that the dVQE algorithm opens up a path to simulate microscopic systems by extending the applicability of the VQE to investigate the intriguing field of open quantum system. 
It is notable that VQE and dVQE algorithms share the strategy to accomplish speed-up over classical algorithms -- optimization of variational quantum circuits to obtain the lowest eigenstate of some Hermitian operator with a polynomial number of Pauli operators. 
We expect to benefit from both when the NISQ devices become powerful enough to simulate ground states of non-local Hamiltonians (e.g. the ones considered in quantum chemistry),
\textcolor{black}{although  it still remains to be an open question whether or not quantum advantage is achievable by our dVQE algorithm.
}

The dVQE algorithm may enable us to explore the wide spectrum of many-body systems that take the environmental effect from, e.g., the solvent, electric voltage, or heat bath into account.
For instance, the dissipative fermionic Hubbard model would be a good test bed for the dVQE algorithm. 
It is known that, under specific quasi-local number-conserving dissipation, this model relaxes into exotic pure states exhibiting $\eta$-pairing order, $d$-wave pairing order, etc.~\cite{diehl_2008, kraus_2008, diehl_2011}.
The stability of such states under perturbation or additional dissipation is believed to be a classically intractable problem which we expect to become feasible via the dVQE algorithm when ground states of the Hubbard model in isolated system are reliably simulated via the VQE algorithm.

We remark several future problems in the following.
First is the approximation procedure of the eigenvalue distribution. 
While the number of measurements required to estimate physical observables remains the identical scaling as in the ordinary VQE simulations, as we have mentioned in Sec.~\ref{subsec:measurement}, it is an intriguing question how the error behaves depending on the eigenvalue distribution. 
\textcolor{black}{Furthermore, it is essential to investigate whether the bottleneck for improving the error is the optimization process or the estimation of eigenvalue distribution.}
Second is to elucidate the performance of Pauli grouping strategy and measurement shot weighing that can be used to suppress the total number of measurement, which determines the computational cost of the algorithm after all.
We raise these points as a future problem together with the development of hardware-friendly ansatz for the eigenvalue distribution.

\section*{Acknowledgements}
We are grateful to Ryusuke Hamazaki for fruitful discussions. 
This work was supported by QunaSys and MEXT Q-LEAP.  
Quantum circuits were numerically simulated with a variational quantum
circuit simulator Qulacs~\cite{qulacs_2018}. 
The actual quantum simulations were performed on the device provided by the Rigetti Quantumm Cloud Service~\cite{rigetti_2016}.

N. Y. was supported by the  JSPS through Program for Leading Graduate Schools (ALPS) and JSPS fellowship (JSPS KAKENHI Grant No. JP17J00743).
K. M. thanks METI and IPA for its support through MITOU Target program. K. M. is also supported by JSPS fellowship (JSPS KAKENHI Grant No. JP19J10978).
K. F. is supported by KAKENHI Grant No. JP16H02211, JST PRESTO JPMJPR1668, JST ERATO JPMJER1601, and JST CREST JPMJCR1673, MEXT Q-LEAP JPMXS01180673.

\appendix

\section{Errors in matrix and vector representations}\label{sec:distance}
\begin{figure}[b]
\begin{center}
\begin{tabular}{c}
  \begin{minipage}{0.97\hsize}
    \begin{center}
     \resizebox{0.9\hsize}{!}{\includegraphics{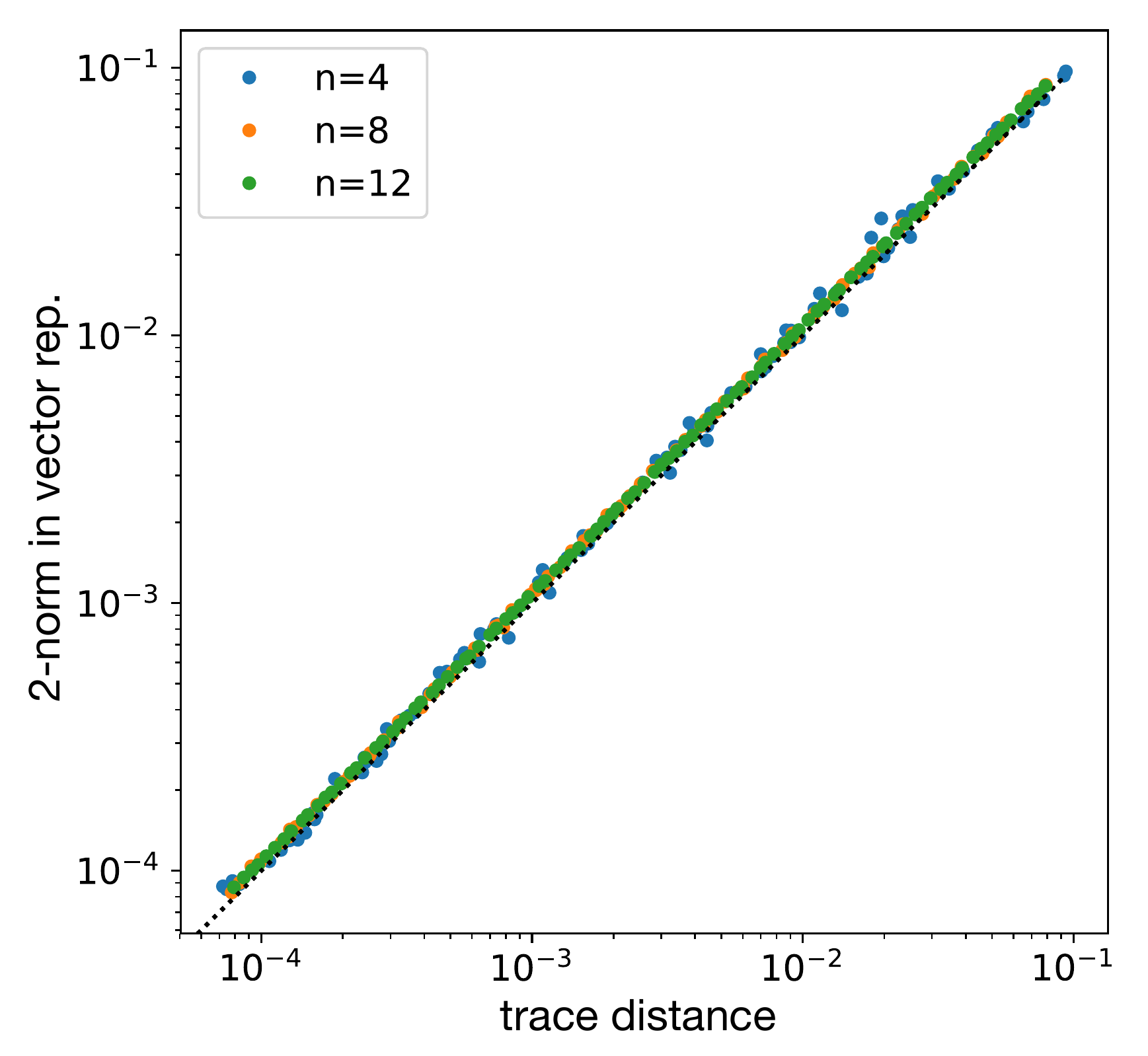}}
   \end{center}
  \end{minipage}
\end{tabular}
\caption{Deviation between trace distance in the matrix representation and 2-norm in the vector representation. Here, 1000 independent data with varying range of noise is displayed.
The blue, orange, and green dots denote data for random mixed states of $n=4, 8, 12$ qubits, and the black dotted line indicates $d_v = d_m$.}
\label{fig:choi_trace_dist}
\end{center}
\end{figure}


\textcolor{black}{The dVQE algorithm employs the vector representation, and therefore the error directly estimated from measurements is related with the 2-norm in the vector representation, i.e., $d_v = \|\ket{\rho}_{\theta} - \ket{\rho_{\rm SS}}\|_2$ where $\ket{\rho_{\theta}}$ and $\ket{\rho_{\rm SS}}$ are the variational ansatz and exact solution, respectively.
On the other hand, the actual error of physical observables are related with the trace distance between two density matrices, $d_m = {\rm Tr}[\sqrt{(\rho_{\theta} - \rho_{\rm SS})^{\dagger}(\rho_{\theta} - \rho_{\rm SS})}]$.
}
\textcolor{black}{While the relationship between $d_v$ and $d_m$ is not evident due to the difference in the normalization, here we give a numerical observation which leads us to expect that values in two measures can be similar to each other.
}

\textcolor{black}{In Fig.~\ref{fig:choi_trace_dist}, we provide the result of a numerical experiment performed to study the difference between $d_v$ and $d_m$.
Each data point is given as follows. First, we generate a random diagonal density matrix $\rho$ whose diagonal elements are drawn randomly from a box distribution on interval $[0,1]$. 
After normalizing $\rho$, we add a small diagonal noise as $\rho' = \rho + \delta$ where each element of $\delta$ is drawn from the normal distribution with various width.
Finally, we calculate both $d_v$ and $d_m$ between $\rho$ and $\rho'$.
Note that both $\rho$ and $\delta$ are generated randomly for each data.
We observe that, while the value of errors varies depending on the width of the normal distribution, $d_v$ and $d_m$ are very close to each other.
We remark that further investigations including rigorous arguments are important future problems.
}

\section{Optimization of variational parameters}\label{sec:seqmin}
\begin{figure*}[t]
    \begin{center}
     \resizebox{0.98\hsize}{!}{\includegraphics{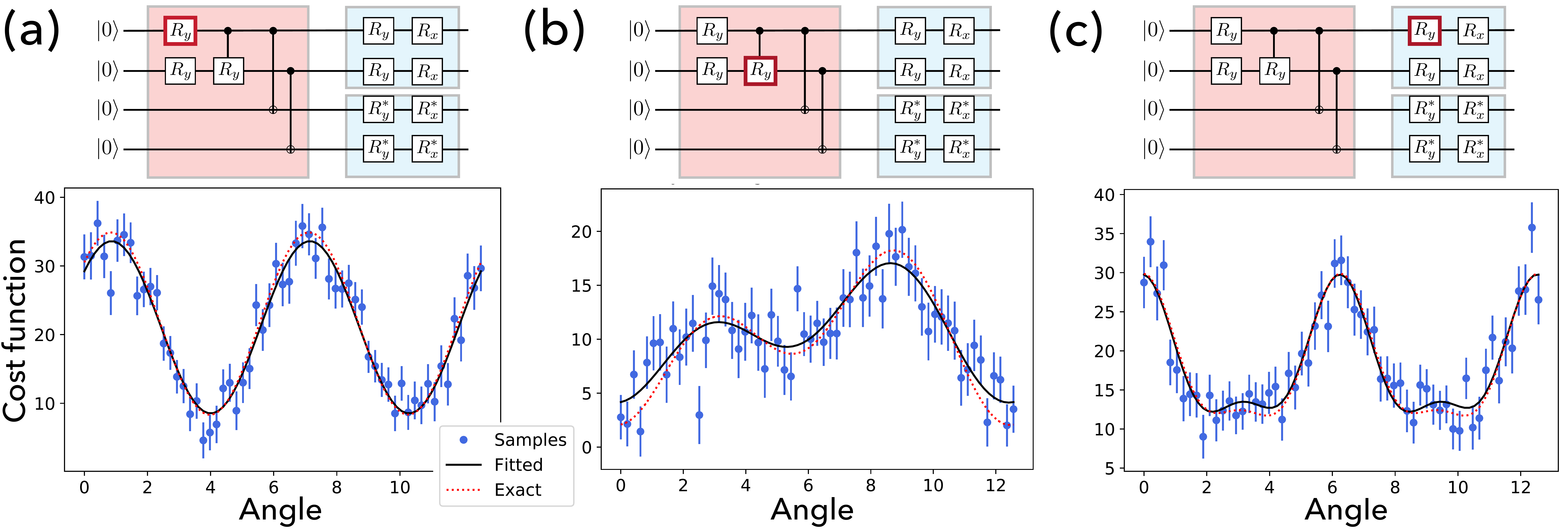}}
   \end{center}
\caption{\label{fig:landscape_cQED_N2_mitigated} (Color Online) Numerical simulation for the landscape of the cost function with respect to single rotational angle for (a) the RY gate in $\widetilde{D}$, (b) the controlled RY gate in $\widetilde{D}$, and (c) the RY gate in $V$. The blue dots are the error mitigated value obtained by sampling from noisy quantum circuits, the black lines obtained from fitting into function with appropriate modes, and the red line is the exact value. The Liouvillian operator is identical to the one employed in Sec.~\ref{subsec:current} with the parameters given as $J=0.3, \lambda=0, \mu = 1, \gamma_1 = 0.3, \gamma_2 = 0.5, \theta = \pi/8$. We take entangled type for the eigenvalue distribution ansatz with the repetition numbers $D_1 = D_2 = 1$. Each Pauli term is sampled over 400 times.
}
\end{figure*}
In this Appendix, we discuss the quantum-classical hybrid optimization of the parametrized quantum circuit for the vector representation of the NESS in open quantum system. 
Let $R_d$ ($R_v$) be the number of parameters for the diagonal matrix $D$ (basis transformation $V$) in the variational quantum circuit that is expressed as
\begin{equation}
    U(\theta) = \left[ V(\theta_v) \otimes V^*(\theta_v) \right] \widetilde{D}(\theta_d),
\end{equation}
where $\theta:= \{\theta_d^r\}_{r=1}^{R_d} \cup \{\theta_v^r\}_{r=1}^{R_v}$ is the set consisting of the total $R=R_d+R_v$ parameters.

As is discussed in Sec.~\ref{sec:method}, the cost function is defined as 
\begin{equation}
    C(\theta) := \braket{0|U^{\dagger}(\theta) \Lind^{\dagger} \Lind U(\theta)|0},
\end{equation}
where $\Lind$ denotes the vector representation of the Lindblad operator.
We may apply the sequential minimal optimization (SMO) technique~\cite{nakanishi_2019}.
This method, explicitly taking advantage of the periodicity of the cost function with respect to the parameters, has been shown to work very efficiently when one has access to the exact expectation values. 
The concrete algorithm of the SMO used in our work is given as follows:
\begin{enumerate}
    \item[1.] Initialize the variational parameters as $\theta_{n=0}$ where $n$ is the number of the update steps already performed.
    \item[2.] Choose the index $r_n\in\{1, ...,R\}$ for the parameter to be optimized. This is done in a sequential manner.
    \item[3.] With the QPU, estimate the $s$-th point of the landscape, $C_n(\theta_{n,s})$, whose parameters are given as
        \begin{eqnarray}
            \theta_{n,s} &=& \{\theta_n^{(1)}, ..., \theta_n^{(r-1)}, \nonumber \\
           && \  s\left(\theta^{(r)}_{\rm max} - \theta^{(r)}_{\rm min}\right)/N_s, \nonumber \\
           &&\ \theta_n^{(r+1)}, ..., \theta_{n}^{(R)}\},
        \end{eqnarray}
        where $\theta_{\rm max}^{(r)}$ and $\theta_{\rm min}^{(r)}$ are the maximum and minimum of the variational parameter determined from the periodicity and restriction. Here, $N_s$ is the number of function evaluation per parameter.
    \item[4.] With the CPU, perform the curve fitting and determine the optimal value $\theta^{(r)}$ that the minimizes the cost function. Replace the previous $\theta_n^{(r)}$ with the optimal value to define the updated parameters $\theta_{n+1}$.
    \item[5.] Repeat 2-4 until the optimization converges.
\end{enumerate}

The periodicity plays important role during the step 4 in the above algorithm.
Shown in Fig.~\ref{fig:landscape_cQED_N2_mitigated} is the comparison between the cost function landscapes calculated exactly and the one estimated from sampled points. 
While the parameters for usual rotational gates exhibit period $2\pi$, as in the panel (a), parameters for controlled rotation gates (panel (b)) or the basis transformation (panel (c)) result in showing multiple modes~\footnote{With $K$ control qubits for the rotational gate, the cost function with respect to its parameter includes terms with period $2\pi, 2^2\pi, \dots, 2^{K+1}\pi$.}.

\section{Mitigation of gate errors}\label{sec:mitigation}
In this Appendix, we introduce the gate error mitigation scheme to obtain the ideal measurement result.

First, for the quantum simulation using the real quantum device provided in Rigetti Quantum Cloud Service, we consider quantum circuits with redundant structure to increase the noise~\cite{heya_2019}. 

Among the most fundamental gate sets employed in the current quantum device, namely $R_z(\phi)$ with arbitrary $\phi$, $R_x(\pm\pi/2)$, and the control $Z$ ($CZ$) gates, the main noise sources are known to be the latter two.
We therefore consider replacing the gates using the following identities for a positive odd integer $\mathcal{E} \in \{1, 3, 5, ...\}$ as,
\begin{eqnarray}
    R_x(\pm \pi/2) &\equiv& R_z(\pi) \left( R_x(\pm \pi/2)\right)^{\mathcal{E}} R_z(\pi),\nonumber \\
    CZ &\equiv& CZ^{\mathcal{E}} \nonumber.
\end{eqnarray}
Since the error for the measurement result with a given $\mathcal{E}$ can be expected to become $\mathcal{E}$-times larger, we apply the linear extrapolation to estimate the result in the clean limit~\cite{temme_2017}.

\begin{figure}[t]
    \begin{center}
     \resizebox{0.98\hsize}{!}{\includegraphics{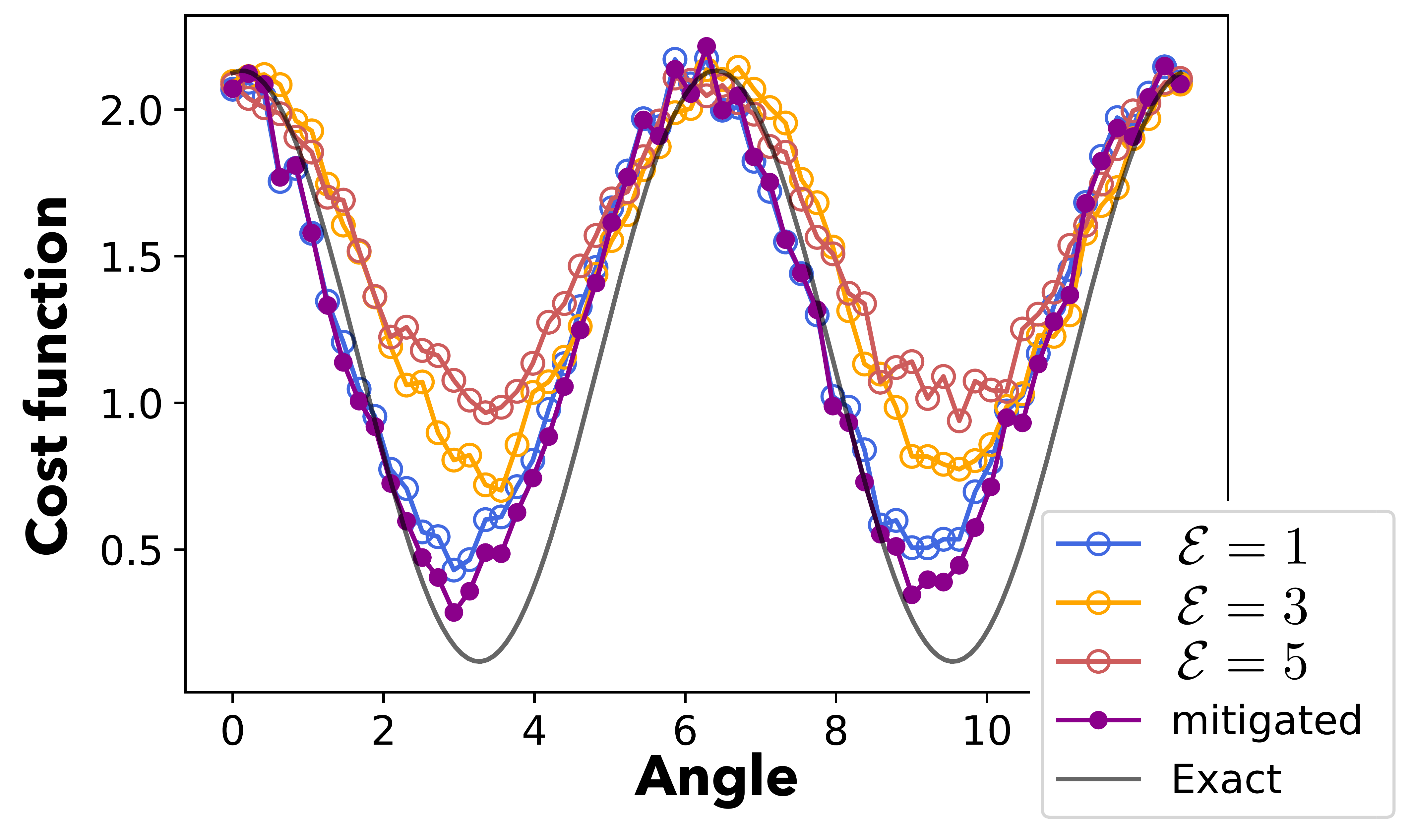}}
   \end{center}
\caption{\label{fig:landscape_TFIM_N1_QPU_mitigated} (Color Online) Quantum simulation of the cost function landscape with gate error mitigation. The blue, yellow, and red unfilled circles are results from a redundant quantum circuit with $\mathcal{E} = 1, 3,$ and $5$, respectively. 
Error mitigated values obtained from linear extrapolation are shown as the purple filled circles, and the exact values are denoted by the real line.
The cost function is defined from the Liouvillian operator identical to the one in Sec.~\ref{subsec:1dTFIM} with the parameters given as $g = 0.5, \gamma_1 = 1, \gamma_2 = 0.5$ with system size $N=1$. We use the decoupled type for the eigenvalue distribution ansatz, and the repetition number for the basis transformation is $D_2=0$. Here, the cost function is computed for the parameter in the eigenvalue distribution, and each Pauli term is sampled over 400 times.
}
\end{figure}

Figure~\ref{fig:landscape_TFIM_N1_QPU_mitigated} shows the comparison of unmitigated and mitigated landscape of the cost function with respect to a specific variational parameter in the ansatz.
Although the error mitigation using three points from $\mathcal{E} \in \{1,3,5\}$ is found to be sufficient to optimize the dissipative Ising model with $N=1$ which use 2 qubits, investigation of larger system size would require improvement of fidelity to estimate the physical observables accurately.

For numerical simulations, in contrast, we consider 
one and two qubit depolarizing noise. Namely, each $k$-qubit gate ($k=1,2$) is assumed to be subject to a linear mapping that operates on a mixed state $\rho$ as
\begin{eqnarray}
    \rho \mapsto \rho' = (1 -p_k)\rho + \frac{4^k-1}{4^k}p_k\mathbbm{1},
\end{eqnarray}
where $p_k$ denotes the error rate.
After measurements under the error rate $p_1 = \mathcal{E}\times10^{-3}$ and $p_2 = \mathcal{E}\times 10^{-2}$ for arbitrary single- and two-qubit gates, respectively, we perform the linear extrapolation. In this paper, we adopted three points by taking $\mathcal{E} \in \{1, 2, 3\}$ as the amplitude of the error.

\section{Numerical simulation without error}\label{sec:exact_simulation}
In the main text, we have provided results of quantum and numerical demonstration of the dVQE algorithm that mitigates the gate error using the technique discussed in Appendix~\ref{sec:mitigation}. 
In the following, we verify the method via noiseless simulation for the quantum Ising model discussed in Sec.~\ref{subsec:1dTFIM}.
The system size is taken as $N=8$ where other model parameters are $\gamma_1 = 1, \gamma_2 = 0.5$ as well as in the main text.
The approximation by the dVQE algorithm results in the infidelity of $\order(10^{-2})$ whose physical observables are in good agreement with the exact solution.

\begin{figure}[t]
\begin{center}
\begin{tabular}{c}
  \begin{minipage}{0.97\hsize}
    \begin{center}
     \resizebox{0.95\hsize}{!}{\includegraphics{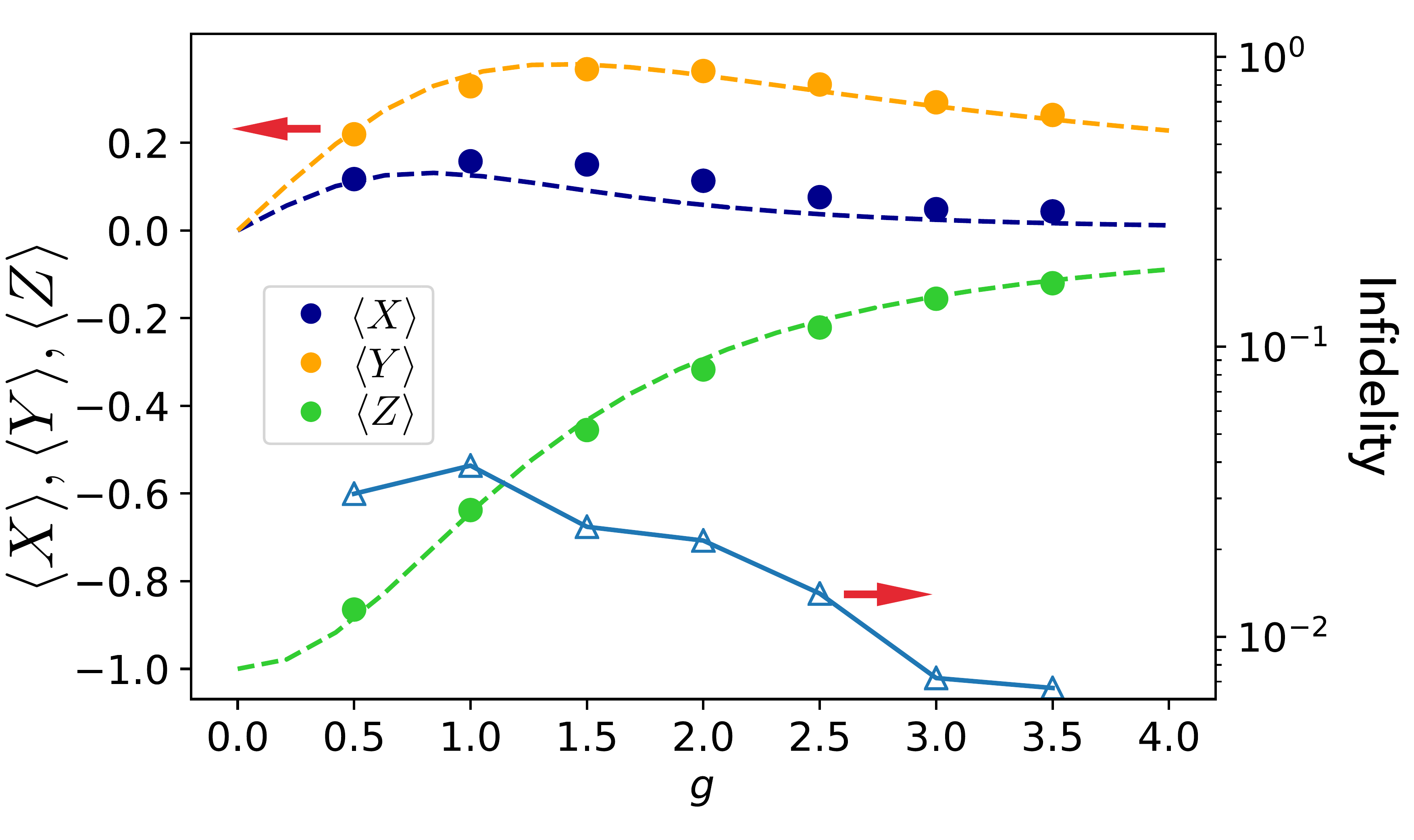}}
   \end{center}
  \end{minipage}
 
\end{tabular}
\end{center}
\caption{\label{fig:observable_1dTFIM_exact} (Color Online) Noiseless numerical simulation of magnetization curves (left axis) for the dissipative quantum Ising model.
The style of the figure is identical to Fig.\ref{fig:observable_1dTFIM_mitigated}. 
Here, the dissipation amplitudes are taken as $\gamma_1 = 1$ and $\gamma_2 = 0.5$ for system size $N=8$. The eigenvalue distribution ansatz is the entangled type shown pictorially in Fig.~\ref{fig:eigval_ansatz}(a). The repetition numbers are taken as $D_1=D_2 = 1$, and hence the circuit consists of 37 parameters in total.
}
\end{figure}

\bibliography{bibliography_open_vqe.bib}
\end{document}